\documentclass[prd, twocolumn, nofootinbib]{revtex4}

\usepackage{epsfig} 

\newcommand{\beq}{\begin{equation}}
\newcommand{\eeq}{\end{equation}}
\newcommand{\beqa}{\begin{eqnarray}}
\newcommand{\eeqa}{\end{eqnarray}}
\newcommand{\om}{\Omega_m}

\def\fun#1#2{\lower3.6pt\vbox{\baselineskip0pt\lineskip.9pt
  \ialign{$\mathsurround=0pt#1\hfil##\hfil$\crcr#2\crcr\sim\crcr}}}

\begin{document} 

\title{Going Nonlinear with Dark Energy Cosmologies} 
\author {Eric V.\ Linder} 
\affiliation{Physics Division, Lawrence Berkeley Laboratory, 
Berkeley, CA 94720} 
\author {Martin White} 
\affiliation{Departments of Physics and of Astronomy, University 
of California, and Physics Division, Lawrence Berkeley Laboratory, 
Berkeley, CA 94720} 

\date{\today} 

\begin{abstract} 
We propose an efficient method for generating high accuracy ($\sim1\%$) 
nonlinear power spectra for grids of dark energy cosmologies. 
Our prescription for matching matter growth automatically matches the
main features of the cosmic microwave background anisotropy power spectrum,
thus naturally including ``CMB priors''.
\end{abstract} 


\maketitle 

\section{Introduction} \label{sec.intro}

The formation of large scale structure in the universe is a central 
element of astrophysics.  Apart from the structure itself, it also 
offers a incisive tool for probing cosmology and particle physics. 
Since structure formation is a competition between expansion and
gravity, the growth as a function of redshift is intimately tied to
the expansion of the universe.  
Studies of matter density fluctuations can thus shed light on the
physics responsible for the accelerating expansion, i.e.\ dark energy.
The growth of large scale structure also provides one of the few tests
of general relativity on cosmological scales.

One of the most informative descriptions of large scale structure, and
the most widely used, is the power spectrum of density fluctuations.
The power spectrum can be probed through observations including galaxy
and cluster redshift surveys, weak gravitational lensing, baryon acoustic 
oscillations, the Lyman alpha forest, etc.
To translate the measured power spectrum to a cosmological model, one
needs accurate predictions for the power spectrum in a variety of dark energy
cosmologies.  This typically comes from large simulations, and fitting
forms derived from such simulations.  However the multitude of possible dark
energy cosmologies is vast, and the necessary simulations are computationally
expensive.  The first dynamical dark energy simulations appeared
only in 2003 \cite{linjen,klypin} and even constant dark energy equation of
state model simulations exist for only a handful of models.
Fitting forms, such as in \cite{pkfit}, are tuned to the cosmological
constant model and even then agree with simulations to only $\sim10\%$.

To use the next generation of large scale structure observations 
we need a better, more efficient method of predicting the effects of 
cosmology on the power spectrum.  Section \ref{sec:linnl} presents a 
new prescription for this, reducing the dimensionality of the grid of 
cosmologies that need simulation.  In \S\ref{sec:sim} we present the 
results of simulations, obtaining $\sim1\%$ accuracy in the matter power
spectra.  The method is extended in \S\ref{sec:wacmb} with a discussion
of time varying equations of state and cosmic microwave background
anisotropies. 

\section{Going Nonlinear \label{sec:linnl}} 

In general relativity the power spectrum of linear matter density
perturbations can be readily calculated (see \cite{SSWZ} for a recent
convergence study).  On large scales and at late times the power spectrum
grows with fixed shape and the growth function can be evaluated, even for
dynamical dark energy cosmologies, from a simple differential equation
\cite{linjen} or highly accurate fitting formulas \cite{groexp}.
This remains true even for some extended theories of gravity \cite{groexp}. 
Since nonlinear structures develop from linear perturbations, the linear
growth function is a natural place to start. 

Given two cosmological models, we can match their growth factors at 
the present by normalizing to the same mass fluctuation 
amplitude, $\sigma_8$.  For definiteness, consider matching some 
dark energy model, characterized by its equation of state or pressure 
to energy density ratio, $w=p/\rho$, to a cosmological constant ($w=-1$) 
cosmology with dimensionless matter density $\om=0.3$ (we assume spatial 
flatness for all models).  Next we match the normalized growth factors 
$\tilde g(a)$ at a second scale factor.  This scale factor, $a_\star$, 
is chosen to 
minimize the deviation between the models' linear growth factors over 
some redshift range.  For the range $z=0-3$, we find $a_\star=0.35$, 
or $z_\star=1.86$, is a good choice.
One could choose a different range and obtain a different $a_\star$.  

Matching the $\tilde g(a_\star)$ determines a value of $\om$ for the 
dark energy model.  Finally, we fix the Hubble constant by keeping 
$\om h^2$ the same for the two models.  Table \ref{tab:wlist} shows 
the results for various dark energy cosmologies.  

The degeneracy 
relation upon changing $w$, or $\om$, between two constant equation 
of state models is 
\beq 
\Delta\om\approx 0.42(1-a_\star)\Delta w. 
\eeq 
For example, if the models differ by 0.02 in $\om$, they must also 
differ by 0.096 in $w$ to match growths at, say, $z=1$ (generalizing 
$a_\star$).  This is only a rule of thumb (e.g.\ a coefficient 0.39 
works better for $w<-1$ models and 0.44 for $w>-1$ models), good 
near the fiducial $\om=0.3$, $w=-1$.  However it is easy
to compute the matching numerically, and we use the exact calculation
throughout.

\begin{center}
\begin{table}[!ht] 
\begin{tabular}{c|c|c|c|c}
$\om$&$h$&$w_0$&$w_a$&$10^5\delta_H$\\ 
\hline 
0.3&0.7&-1&0&5.0\\ 
0.26&0.7519&-1.156&0&5.769\\ 
0.28&0.7246&-1.076&0&5.357\\ 
0.32&0.6778&-0.929&0&4.688\\ 
0.34&0.6575&-0.861&0&4.412\\
0.3846&0.6182&-0.8&0.3&3.900\\ 
0.3356&0.6618&-0.8&-0.3&4.470\\
0.3214&0.6763&-1.0&0.3&4.667\\ 
0.2808&0.7235&-1.0&-0.3&5.342\\ 
\hline 
\end{tabular} 
\caption{Cosmological parameters generated from our prescription 
for matching the nonlinear power spectra of dark energy cosmologies. 
The last column is defined as the present day value of the dimensionless 
power spectrum at $k=H$, i.e.\ $\Delta^2(k)=\delta_H^2 (k/H)^{3+n}T^2(k)$, 
where we adopt scale invariance $n=1$, and $T(k)$ is the transfer function 
\cite{EisHu}, fixed for $k$ measured in 1/Mpc.}  
\label{tab:wlist}
\end{table} 
\end{center}

\section{Results from Simulations \label{sec:sim}}

By working at constant $\Omega_mh^2$ and $\Omega_bh^2$ we fix the shape
of the matter power spectrum in physical units, e.g.~$k$ in Mpc${}^{-1}$. 
Our procedure arranges a close match of the growth of the power spectrum 
in the linear regime.
Thus it is natural to expect that the nonlinear power spectra will be
closely matched\footnote{This is certainly true to second order in
perturbation theory \cite{PT} -- see \cite{PTrev} for a review.}.
To test this we make use of N-body simulations.

For the first 5 models of Table \ref{tab:wlist} we ran N-body simulations
using a parallel particle-mesh (PM) code \cite{PM}.
In each case we evolved $256^3$ equal mass particles using a $512^3$ force
mesh from $z=60$ to the present, dumping the phase space data at $z=1.86$, 
1, and 0. 
The simulation volume for the results presented here was a periodic, cubical
box of side 366 Mpc, though we also ran a few larger boxes to check
convergence on the largest scales.
For each output time, the power spectrum was computed by assigning the
particles to the nearest grid point of a regular, $512^3$ Cartesian mesh
and Fourier transforming the resultant density field.
The resulting $\Delta^2(k)$, corrected for the assignment to the grid using
the appropriate window function and for Poisson shot noise, were placed in
logarithmically spaced bins of $k$.
The average $\Delta^2(k)$ is plotted at the position of the average $k$ in
each bin.

\begin{figure}
\begin{center}
\psfig{file=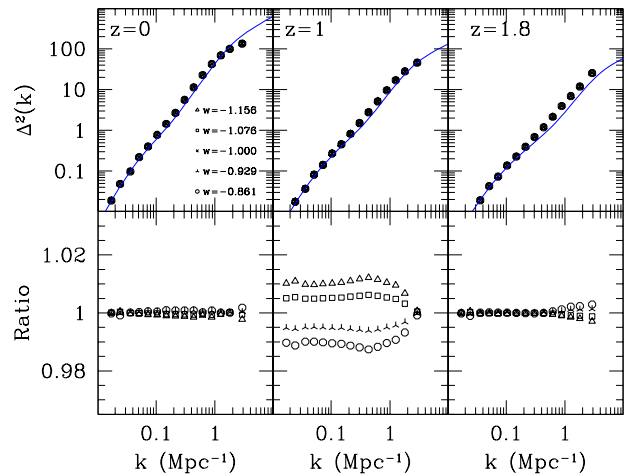,width=3.4in}
\caption{The nonlinear mass power spectrum is shown at three redshifts. 
The top panels plot $\Delta^2(k)=k^3P_k/(2\pi^2)$ from the simulations 
for five different dark 
energy models matched according to our prescription.  The solid line 
indicates the Smith et al.\ \cite{pkfit} fitting formula.  The bottom 
panels show the ratio of $\Delta^2(k)$ for the four dark energy models 
relative to the $\Lambda$CDM case. 
}
\label{fig:pk} 
\end{center}
\end{figure}

While the PM code has limited force resolution, it allows high mass
resolution at modest computational cost.  This is important in the
translinear scales ($\Delta^2\sim 1$) of interest here.
The finite force and mass resolution of the PM code limits the dynamic
range in $k$ to $\sim 50$, and the short-fall in power can be easily
seen in the highest $k$ point plotted in the top left panel of
Fig.~\ref{fig:pk}.  Much of the effect of the force resolution is reduced in
taking the ratio (see also \cite{McDonald}) of power spectra, as in the
lower panels of Fig.~\ref{fig:pk}, although the details of the shot-noise
correction are starting to become important on these scales.

{}From Fig.~\ref{fig:pk} we see that across most of the range of scales
probed by, e.g., weak lensing\footnote{For sources at $z_s\sim 1$, or comoving
distance $3\,$Gpc, wavenumber $k=1\,{\rm Mpc}^{-1}$ corresponds to
multipole $\ell\sim k\chi=1500$, or arcminute scales, at the peak of the
lensing kernel, $\chi=1.5\,$Gpc.} and for which we 
believe purely gravitational N-body 
simulations will provide accurate calculations of the power spectrum, 
the models are indeed nearly degenerate.  The maximum deviation in 
$\Delta^2(k)$ 
is 1.5\%, for the $\om=0.26$, $w=-1.156$ model at $z=0.5$.  This is also
the maximum deviation in the (square of the) linear theory growth rates.  
Where the 
growth rates are closely matched, the nonlinear power spectra are too.

\section{Growth and the CMB \label{sec:wacmb}} 

A significant new result is that the distance to last scattering $d_{\rm lss}$ 
is automatically preserved through our matching prescription.  That is, 
by matching the growth today and at $a_\star=0.35$, we ensure that 
$d_{\rm lss}$ changes very little; the growth factor determines the distance. 
The fractional distance deviation of the dark energy models from the 
$\Lambda$CDM case is less than $0.04\%$ (e.g.\ 0.001\% for the $\om=0.28$, 
$w=-1.076$ case).  

This is important because the CMB is one of our
strongest probes of large scale structure and cosmology.
The main impact of the dark energy and the matter density on the CMB 
power spectrum is through the geometric degeneracy involving the 
distance to the last scattering surface, and the physical matter 
density $\om h^2$.  Our prescription takes care of both, meaning that
we select models that automatically include a strong CMB prior
(see also \cite{LensGrid}): our matched models, with $\om h^2$ and
$d_{\rm lss}$ fixed, will agree on the CMB temperature power spectrum
to high accuracy, except at low multipoles where the integrated Sachs-Wolfe
effect enters (though cosmic variance makes differences hard to detect)
and at high multipoles through the effect of gravitational lensing.

We can further extend this to models with time varying dark energy 
equation of state.  For values of $w_0$ and $w_a$, where 
$w(a)=w_0+w_a(1-a)$, we match the growth at the present and at $a_\star$ 
(we have checked that $a_\star=0.35$ works for these cases as well). 
This provides the necessary value of $\om$, and hence $h$ (see 
Table \ref{tab:wlist}).  For matching models with $(w_0,w_a)$ of $(-1,0.3)$, 
$(-1,-0.3)$, and $(-0.8,-0.3)$, the maximum deviation from the fiducial 
$\Lambda$CDM model in the linear growth factor is less than 0.6\%, at 
$z=0.6$.  So we expect, based on \S\ref{sec:sim}, that the nonlinear 
power spectra will match to within 1.2\%.  A matching model with 
$(-0.8,0.3)$, where the average equation of state lies further from 
the fiducial $w=-1$ and the required $\om$ is 30\% different from the 
fiducial model, fits somewhat less well: the expected deviation 
in the nonlinear power spectrum is 3.4\%. 

The time varying equation of state models still preserve the CMB 
matching as well.  They match on $d_{\rm lss}$ to better than 0.1\%, 
except for $(-0.8,0.3)$ which agrees to 0.4\%. 

\section{Conclusion} 

We present a simple prescription for generating the nonlinear mass 
power spectrum of dark energy cosmologies, accurate to $\sim1-2\%$. 
It requires only straightforward calculation of the linear growth 
factor at two redshifts.  This will allow efficient generation of 
a suite of dark energy cosmologies using only a reduced dimension 
grid of $\Lambda$CDM simulations. 

Moreover, we have identified a strong 
degeneracy between the growth factor and the CMB distance to last 
scattering, such that our matching procedure can automatically 
satisfy CMB constraints.

\begin{acknowledgments} 
This work has been supported in part by the Director, Office of Science,
Department of Energy under grant DE-AC02-05CH11231, by the NSF and by
NASA.
Some of these simulations were run on the IBM-SP at NERSC.
\end{acknowledgments}

\end{document}